\documentclass[twocolumn,floatfix,preprintnumbers,amsmath,amssymb,prb]{revtex4}

\usepackage{graphicx}
\usepackage{dcolumn}
\usepackage{bm}

\begin{document}


\title{Shake-up Processes in a Low-Density Two-Dimensional Electron Gas:
Spin-Dependent Transitions to Higher Hole Landau Levels}
 \author{A. B. Dzyubenko}
\affiliation{Department of Physics,
California State University at Bakersfield, Bakersfield, CA 93311, USA \\
Department of Physics, University at Buffalo, SUNY, Buffalo, NY 14260, NY}
\email{adzyubenko@csub.edu}
\altaffiliation[]{on leave from General Physics Institute, RAS,
                Moscow 117942, Russia.}

\date{\today}

\begin{abstract}
A theory of shake-up processes in photoabsorption of an
interacting low-density two-dimensional electron gas
(2DEG) in strong magnetic fields is presented.
In these processes, an incident photon creates an electron-hole pair and,
because of Coulomb interactions, simultaneously
excites one particle to higher Landau levels (LL's).
In this work, the spectra of correlated charged spin-singlet and spin-triplet
electron-hole states in the first hole LL and
optical transitions to these states (i.e., shake-ups to the first hole
LL) are studied. Our results indicate, in particular, the
presence of optically-active three-particle quasi-discrete
states in the exciton continuum that may give rise to
surprisingly sharp Fano resonances in strong magnetic fields.
The relation between shake-ups in photoabsorption of the 2DEG and
in the 2D hole gas (2DHG), and shake-ups of isolated negative
$X^-$ and positive $X^+$ trions are discussed.
\end{abstract}



\maketitle

\section{\label{sec:level1} Introduction }

Photoluminescence (PL) and photoluminescence excitation (PLE) spectroscopy
of a two-dimensional electron gas (2DEG) in a perpendicular magnetic field
proved effective in studying few- and many-body correlation effects
(see, e.g.,
Refs.~\onlinecite{Kukushkin,Fink96,Yak97,Ossau2001,trions,Yusa2001,Schueller2002,Broocks2002,PRL2002,Sanvitto}
and references therein).
In the 2DEG of density $n_e$ in quantizing magnetic fields
with electron filling
factor $\nu_e = 2\pi l_B^2 n_e  < 1$,
the lowest energy absorption channel corresponds to
creation of a bound $e$-$h$ pair, a neutral magnetoexciton
in zero Landau levels (LL's), ${\sl photon} \rightarrow X_{00}$;
$l_B=(\hbar c/eB)^{1/2}$ is the magnetic length.
Transitions to higher LL's,  ${\sl photon} \rightarrow X_{N_eN_h}$,
are also allowed when $N_e=N_h$. All these transitions are strong
and gain strength with increasing magnetic field $B$.
With increasing electron density $n_e$, these transitions
become broadened by exciton-electron interactions,
and optical processes associated with {\em three-particle\/}
bound $2e$-$h$ states of charged excitons, or trions,
$X^-$, start to dominate the absorption \cite{Yak97,Ossau2001}
and PL \cite{trions} spectra.
Correlation effects in electron LL's manifest themselves
in the rich PL spectrum in strong magnetic fields.
\cite{Yusa2001,Schueller2002,Broocks2002,PRL2002,Sanvitto,Ashkinadze}
The various theoretical approaches were used for studying
three-particle bound trion $X^-$ states
\cite{Palacios96,Whit97,PRL2000,Wojs2000,Riva2001,PRB2002}
and magneto-PL of the 2DEG \cite{Cooper97,Hawrylak97,Rashba2000}
in strong magnetic fields.
Despite a large amount of experimental and theoretical work,
the spectra are not fully understood even in the low-density 2DEG
limit. The role of electron-electron interactions, \cite{Sanvitto} disorder
and localization, \cite{PRL2000} and the interplay between
them \cite{Ashkinadze} are yet to be fully assessed.

Other optical manifestation of many-body effects
in the interacting 2DEG are shake-up processes
\cite{Fink96} in magneto-PL: After the recombination of an
$e$-$h$ pair, one electron is excited to one of the higher
LL's. A closely related phenomenon in absorption (PLE)
was also identified \cite{Yak97,Ossau2001}
in low-density 2DEG systems: Here, an incident photon creates an
$e$-$h$ pair and simultaneously excites one electron to
higher LL's. This process was called\cite{Yak97}
``combined exciton-cyclotron resonance''
(ExCR) and may be considered to be a shake-up process
(or a radiative Auger process) in magneto-photoabsorption of the 2DEG.
The shake-up phenomena in absorption and magneto-PL and the relation between
them remain only partially understood. \cite{Fink96,Yak97,Ossau2001,PRB2001}

At finite filling factors $\nu_e$,
magneto-optical absorption involves many-particle
``$N$-electron''  initial states and
``$(N+1)$ electrons + one valence band hole''
final states. Generally,  one has to deal with a many-body problem that is
difficult to solve because the 2DEG in lowest LL's is a strongly
correlated system with no obvious small parameters.
In the low-density limit, however,
when  filling factor $\nu_e = 2\pi l_B^2 n_e \ll 1$,
distance between electrons is large (compared to $l_B$),
and  absorption can be considered to be an optical resonance
$e^-_0 + {\sl photon} \rightarrow 2e$-$h$,
which involves an interacting charged three-particle system of two
electrons and one hole in the final state. \cite{Yak97,Ossau2001,PRB2001}
A similar situation occurs in the vicinity of integer filling
factors. \cite{Cooper97,Hawrylak97,Rashba2000,PRL2002}
In this limit, one needs
(i) to study the spectra of charged three-particle Coulomb $2e$-$h$
problem in a magnetic field, and (ii) to find selection rules and
dipole transition matrix elements.
The techniques \cite{PRL2000,PRB2002,JETPL99}
exploiting an exact symmetry, magnetic translations, \cite{magS}
were applied in our previous paper \cite{PRB2001}
to describe the spectra of charged $e$-$h$ states in higher
LL's and ExCR transitions in the high-field and low-density limit.
In particular, a double-peak structure for the transitions to
the first electron LL was predicted. \cite{PRB2001}
The origin of the second peak in
ExCR are transitions to the continuum formed by optically
{\em inactive\/} excitons $X_{N_eN_h}$ ($N_e \neq N_h$)
which, somewhat unexpectedly,
may have large ExCR oscillator strengths.
Some experimental indications
of the double-peak structure of the ExCR transitions to the first
electron LL  were reported recently. \cite{Ossau2001}

In this work, which is an extension of Ref.~\onlinecite{PRB2001},
we study magneto-optical absorption in the low-density 2DEG in which the
hole---in the presence of excess electrons---is excited
to higher hole LL's. In what follows,
such shake-up absorption processes will be called hole-ExCR.
The theory developed in
Sect.~\ref{sec:theory} and Sect.~\ref{sec:results}
of this paper predicts the shape and
fine structure of the {\em high-energy\/} tail of the main
magneto-absorption peak of the 2DEG.
No detailed experimental studies of this energy region
(with sufficiently high spectral resolution) are known
to the author.

The results of the present theory have also some relevance
for the studies of a two-dimensional hole gas (2DHG).
Experimentally,  Coulomb correlation effects in magneto-PL
spectra of the 2DHG were studied in a number of
papers. \cite{Butov94,Kulik95,Ponomarev96,Volkov97,GlasbergPRB2001}
Note that 2DHG magneto-PL spectra are
often featureless and characterized by broad
lines associated with the Zeeman split levels.
This is mainly because of a large difference in
electron and hole effective masses.
For example,  in GaAs quantum wells the electron and hole cyclotron
energies are $\hbar\omega_{\rm ce} = 1.7$~meV/T and
$\hbar\omega_{\rm ch} = 0.2$~meV/T,
respectively, and differ by nearly an order of
magnitude. \cite{GlasbergPRB2001}
Recently, however, by examining the {\em low energy\/}
tail of the magneto-PL spectra of the
low-density 2DHG Glasberg {\it et al.\/} \cite{GlasbergPRB2001}
succeeded in resolving several groups of recombination
lines. Each group consists of several equidistant peaks
directly related to the valence band LL's (see also
Refs.~\onlinecite{Butov94,Kulik95,Ponomarev96,Volkov97}).
In particular, in the dilute 2DHG the shake-up recombination
lines of the positively charged exciton $X^+$, a complex consisting
of two holes and one electron ($2h$-$e$), were observed.
Quantitatively, the present theory cannot be directly applied
for a description of the 2DHG magneto-optics.
However, as is shown in Sec.~\ref{sec:SU},
there are qualitative similarities between the 2DHG and 2DEG
magneto-absorption spectra in the strong magnetic field and
low-density limit.  We also discuss shake-ups in magneto-PL
of positive $X^+$ and negative $X^-$ trions in Sec.~\ref{sec:X-}.
A short summary is given in Sec.~\ref{sec:summary}.

\section{\label{sec:theory} Theory }

The paper examines the hole-ExCR transitions
$e^-_0+ {\sl photon} \rightarrow 2e$-$h$
in the spin-polarized 2DEG in zero electron  LL $N_e=0$\,$\uparrow$
with final three-particle states $2e$-$h$ that belong to
the first {\em hole\/} LL $N_h=1$
with electrons still in zero LL's (see Figs.~1,\,2).
The classification of states according to electron and hole
LL's is valid in sufficiently strong magnetic fields $B$ such that
\begin{equation}
      \label{highB}
 \hbar\omega_{\rm ce} \, , \,  \hbar\omega_{\rm ch} >
 E_0 = \sqrt{\frac{\pi}{2}} \, \frac{e^2}{\epsilon l_B} \, ,
\end{equation}
where $E_0$ is the characteristic energy of the Coulomb interactions.
Correlated electron-hole states can be obtained in this regime as
an expansion in electron and hole LL's.
Before describing  this procedure, some general  features of the spectra
that hold at arbitrary fields are discussed below.

\begin{figure}[t]
\includegraphics[scale=0.43]{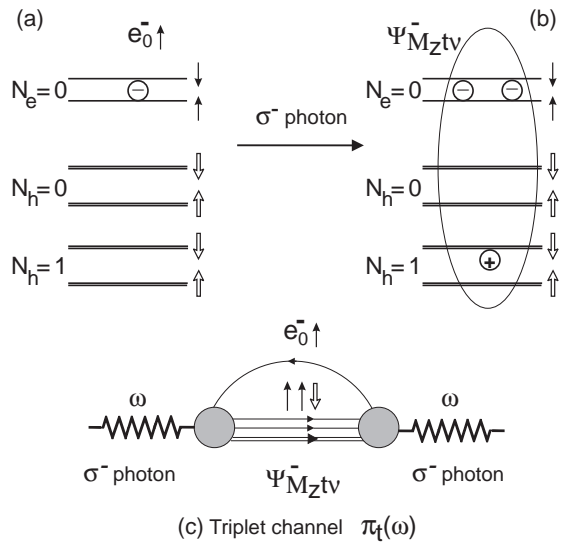}
\caption{  \label{fig:triplet}
Initial  (a) and final (b) states in hole-ExCR in the dilute spin-polarized
2DEG with electron filling factor $\nu_e = 2\pi n_e l_B^2 \ll 1$
in $\sigma^{-}$ polarization. The exact selection rule is $M_z=0$.
(c) Polarization operator $\pi_t(\omega)$ describing
interband transitions.
 }
\end{figure}

Charged $2e$-$h$ states form families of degenerate states;
each family is labeled by the index $\nu$ that plays a role
of the principal quantum number and can be discrete
(bound $X^-$ states) or continuous
(unbound $X+e^-$ states forming a continuum). \cite{PRL2000}
There is a macroscopic number of degenerate states in each family
labeled by the discrete exact oscillator quantum number $k$
($k=0, 1, \ldots$).  This quantum number
is associated with magnetic translations \cite{magS,PRL2000,PRB2002}
and describes the position of the center of the ``cyclotron''
orbit of the composite charged complex in $B$.
Each family starts with its Parent State
$\Psi^-_{k=0 \, M_z \, S_e S_h \nu}$ that has $k=0$ and,
roughly speaking, rotates in a magnetic field about the origin.
All other daughter states,
$\Psi^-_{k\,M_z-k\,S_eS_h\nu}$
(with $k=1,2,\ldots$)
can be constructed iteratively \cite{PRL2000}
out of the Parent State; note the $M_z-k$ values.
They have exactly the same physical properties and, because
of this, will mostly be neglected in the discussion that follows.
Accordingly,  the quantum number $k=0$ of the Parent States will
also be suppressed in notations.
Importantly, the Parent State has the largest possible in the family
value of the total angular momentum projection $M_z$.
It is this value that enters  \cite{JETPL99,PRL2000}
into the exact optical selection, see Eqs.~(\ref{D}) and (\ref{X-SU}) below.
In addition to orbital quantum numbers $k$ and $M_z$,
the three-particle states are characterized by the total spin of two electrons,
either $S_e=0$ (electron spin-singlet states, $s$) or $S_e=1$
(electron spin-triplet states, $t$)
and by the spin state of the heavy hole $S_{h}=\pm 3/2$ ($\Uparrow$,
$\Downarrow$).
The heavy holes are described in this work in the
effective mass approximation and the states of light holes are neglected.

\begin{figure}[t]
\includegraphics[scale=0.43]{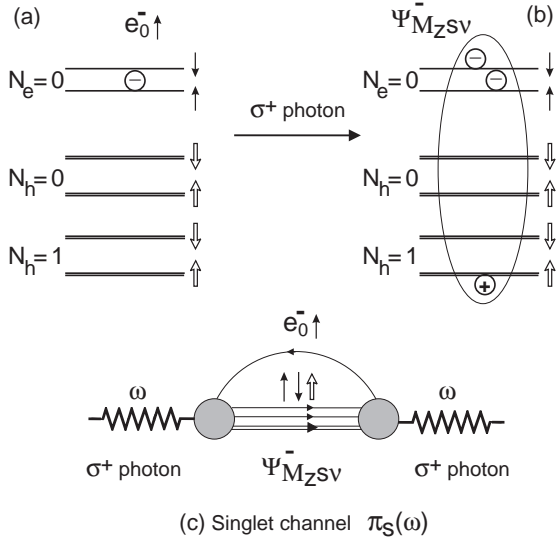}
\caption{\label{fig:singlet}
Same as in Fig.~\protect{\ref{fig:triplet}} for
$\sigma^{+}$ polarization leading to final
electron singlet states $\Psi^-_{M_zs\nu}$.
}
\end{figure}

The exact magneto-optical selection rules, in particular for ExCR,
can be derived  \cite{PRB2001} as follows:
Interband transitions with $e$-$h$ pair creation
are described by the luminescence operator
$\hat{\cal L}_{\rm PL}= p_{\rm cv} \int \! d\bm{r} \,
\hat{\Psi}^{\dagger}_{e}(\bm{r}) \hat{\Psi}^{\dagger}_{h}(\bm{r})$,
where $p_{\rm cv}$ is the interband momentum matrix element.
When an $e$-$h$ pair is photocreated in the presence
of the low-density 2DEG in $N_e=0$ LL,
the dipole transition matrix element to the $\nu$-th family
can be written as
\begin{equation}
       \label{ExCR}
   D(\nu) = \langle \Psi^-_{M_z \, S_e S_h \nu}| \hat{\cal L}_{\rm PL}
              |e^{-}_{0} \rangle  \, ,
\end{equation}
where  only the final state three-particle correlations are taken into account.
The combination of the two exact dipole selection rules,
(i) conservation of the oscillator quantum number,
$\Delta k = 0$ (the centers-of-rotation of charged complexes in the
initial and final states are conserved) and
(ii) no change in the total angular momentum
for the envelope functions, $\Delta M_z= 0$, leads to a very simple but
powerful result:
\begin{equation}
       \label{D}
         D({\nu}) \sim \delta_{N_e=0,M_z} \, ,
\end{equation}
where $M_z$ is the angular momentum projection
of the Parent State in the $\nu$-th family. \cite{PRL2000,PRL2002}
The meaning of Eq.~(\ref{D}) is that
in the magneto-absorption processes involving electrons from zero LL  $N_e=0$,
the achievable final $2e$--$h$ states must have $M_z=0$.
If the 2DEG is spin-up polarized ($\uparrow$, $S_{\rm ez}=1/2$), the
photons of $\sigma^-$ and $\sigma^+$ circular polarizations produce
the electron triplet $t$
and singlet $s$ final states,
respectively (Figs.~\ref{fig:triplet}, \ref{fig:singlet}).
The spin-dependence follows from the form of the
luminescence operator in two circular polarizations,
\begin{eqnarray}
       \hat{\cal L}_{\rm PL}(\sigma^+) & = &
         p_{\rm cv} \int \! d\bm{r} \,
          \hat{\Psi}^{\dagger}_{e\downarrow}(\bm{r})
          \hat{\Psi}^{\dagger}_{h\Uparrow}(\bm{r}) \, , \\
       \hat{\cal L}_{\rm PL}(\sigma^-) & = &
         p_{\rm cv} \int \! d\bm{r} \,
          \hat{\Psi}^{\dagger}_{e\uparrow}(\bm{r})
          \hat{\Psi}^{\dagger}_{h\Downarrow}(\bm{r}) \, .
\end{eqnarray}
Interband transitions are described by
polarization operators $\pi_t(\omega)$ and $\pi_s(\omega)$
that involve a single-particle electron propagator in $N_e=0$\,$\uparrow$ LL
and three-particle propagators, which describe correlated
negatively charged states $\Psi^-_{M_z \, S_e S_h \nu}$ with $M_z=0$
(see Figs.~\ref{fig:triplet}c,\,\ref{fig:singlet}c).
The electron-photon vertices, $D$,  are determined by
Eq.~(\ref{ExCR}), and absorption at frequency $\omega$ is given by
$-2{\rm Im}\,\pi_{t(s)}(\omega) \propto |D|^2 \nu_e $,
where the dependence on electron filling factor $\nu_e$ comes from the
electron propagator.

Let us now turn to the high-field regime (\ref{highB})
in the strictly-2D geometry.
The final three-particle $2e$-$h$ states in ($N_eN_h$)  electron and
hole LL's are found using an expansion in the
complete orthonormal basis of squeezed oscillator
states \cite{PRB2002} which simultaneously preserves both
axial and magnetic translational symmetries
\begin{equation}
        \label{basis}
  |N_R N_r N_h ; k \, m \, l \rangle  \, .
\end{equation}
The conserved oscillator quantum number is fixed automatically in
basis (\ref{basis}) and equals
$k$, while the total angular momentum projection must be
fixed by imposing the condition $M_z = N_R + N_r - N_h - k - m + l$.
Here $N_h$ is the hole LL number, $N_r$ and $N_R$ are the LL numbers of
the electrons relative and center-of-mass motions,
respectively ($N_r + N_R=N_e$).
The electron permutational symmetry requires that the
two-electron relative angular momentum projection
$m_z = N_r-m$ must be even (odd) for electron singlet
$S_e=0$ (triplet $S_e=1$) states;
see Ref.~\onlinecite{PRB2002} for more details.
The quantum numbers $k$, $M_z$, $S_e$, and $S_h$ are
exact so that the total Hamiltonian is block-diagonal in them.
A weak exchange $e$-$h$ Coulomb interaction, which mixes different
electron and hole spin states, is neglected in this work.

Because of the degeneracy in $k$, in what follows only
Parent States $ \Psi^-_{M_z s(t) \nu} $ are considered. The
 quantum numbers  $k=0$ and $S_h$ are omitted for brevity
and $s$ ($t$) denotes the singlet $S_e=0$ (triplet $S_e=1$)
electron spin state.
Neglecting mixing between LL's (the high-field limit),
the triplet $2e$-$h$ states in zero electron and first hole
($N_eN_h$)=(01) LL's,
$\Psi_{M_z t \nu}^{(01)-}$, can be obtained as
the following  expansion in (\ref{basis})
\begin{equation}
        \label{Psi01}
 \Psi_{M_z t \nu}^{(01)-}   =
  \sum_{l=0}^{\infty} \sum_{m=0}^{\infty}
   \alpha_{M_zt\nu}(2m,l) \, |0 \, 0 \, 1 ; 0 \,\, 2m \,\, l \rangle \, ,
\end{equation}
where expansion coefficients
$\alpha_{M_z t \nu}(2m,l) \sim \delta_{M_z,l-2m-1}$
reflect the block-diagonal structure of the Hamiltonian.
These are found by solving the corresponding secular equation.
The Coulomb matrix elements of the total Hamiltonian
are calculated analytically \cite{PRB2002} and the
eigenspectra are obtained by numerical diagonalization of
finite matrices of order $5 \times 10^2$.
Such finite-size calculations do not break either
magnetic translational or axial symmetries and provide very high
accuracy for bound $X^-$ states.
It is important to stress that such calculations are also capable of
reproducing the structure of the
three-particle continuum in some detail. \cite{PRB2002}
The singlet states $\Psi_{M_z s \nu}^{(01)-}$ are
found using the expansion
\begin{equation}
        \label{Psi01s}
 \Psi_{M_z s \nu}^{(01)-}    =
  \sum_{l=0}^{\infty} \sum_{m=0}^{\infty}
   \beta_{M_zt\nu}(2m\!+\!1,l) \, |0 \, 0 \, 1 ; 0 \,\, 2m\!+\!1 \,\, l \rangle
\end{equation}
similar to (\ref{Psi01}) with expansion coefficients
$\beta_{M_z s \nu}(2m\!+\!1,l) \sim \delta_{M_z,l-2m-2}$.

\begin{figure}[t]
\includegraphics[angle=-90,scale=0.39]{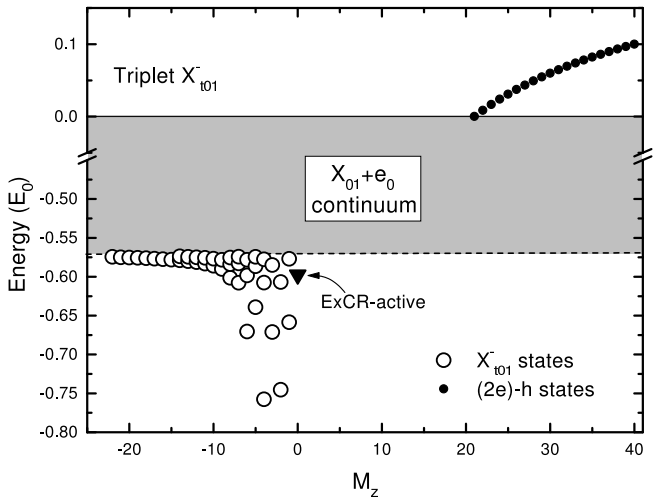}
\caption{          \label{fig:t-levels}
Three-particle electron triplet states $\Psi_{M_z t \nu}^{(01)-}$
in ($N_eN_h$)=(01) LL's.
Open dots correspond to low-lying bound states, charged magnetoexcitons
$X^-_{t01}$. The filled triangle depicts the ExCR-active $X^-_{t01}$ state
with $M_z=0$. Filled dots are excited states in which the hole is
bound to the two-electron pair.
}
\end{figure}

The ExCR transitions are absent in the high-field limit (when no
LL mixing allowed). \cite{PRB2001}
The lowest order non-vanishing contribution to the dipole transition
matrix element (\ref{ExCR}) appears when
the triplet $2e$-$h$ states in {\em zero\/} LL's
are admixed to (\ref{Psi01}):
\begin{equation}
        \label{exp2}
\delta \Psi_{M_z t \nu}^{(01)-}  =
 \sum_{l=0}^{\infty} \sum_{m=0}^{\infty}
  \gamma_{M_zt\nu}(2m,l) \, |0 \, 0 \, 0 ; 0 \,\, 2m \,\, l \rangle \, .
\end{equation}
The expansion coefficients
$\gamma_{M_zt\nu}(2m, l) \sim \delta_{M_z,l-2m}$
are non-zero only because of the Coulomb mixing of different
LL's. Therefore, their characteristic smallness
in high fields is
$\gamma_{M_zt\nu} \sim E_0/\hbar\omega_{\rm ch} \sim B^{-1/2}$.
A similar procedure is used to find
$\delta\Psi_{M_z s \nu}^{(01)}$ for the singlet states.
Technical details of calculations of the dipole
transition matrix elements have been described
elsewhere. \cite{PRB2001}

\section{\label{sec:results} Results and discussion }

\begin{figure}[t]
\includegraphics[angle=-90,scale=0.39]{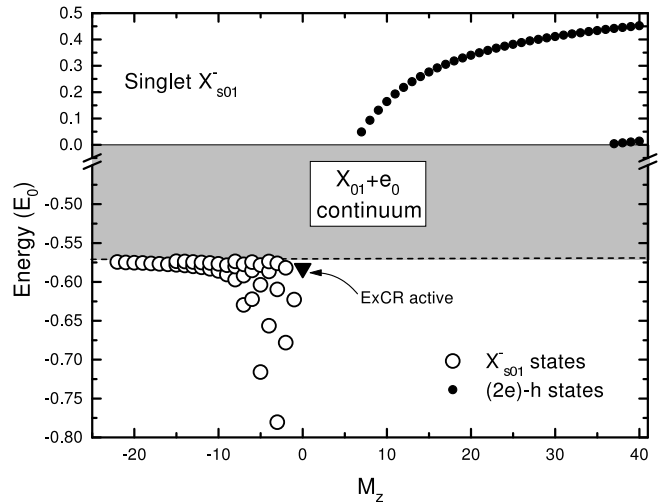}
\caption{         \label{fig:s-levels}
Same as in Fig.~\protect\ref{fig:t-levels} for the  electron singlet states
$\Psi^{(01)-}_{M_zs\nu}$.
}
\end{figure}

The triplet $ \Psi_{M_z t \nu}^{(01)-} $ and singlet
$\Psi_{M_z s \nu}^{(01)-}$ eigenspectra in the
($N_eN_h$)=(01) LL's
are presented in Fig.~\ref{fig:t-levels} and Fig.~\ref{fig:s-levels},
respectively.
The eigenstates were calculated in the strictly-2D high-field limit and
their Coulomb energies are shown relative to free
LL's, $\hbar(\omega_{\rm ce} + 3\omega_{\rm ch})/2$.
The shaded areas of width $0.574E_0$ in
Figs.~\ref{fig:t-levels}, \ref{fig:s-levels}
correspond to the three-particle continuum formed by
the neutral magnetoexciton $X_{01}$
($e$ in its zero and the hole $h$ in the first LL)
with the second electron in a scattering state in zero LL\@.
The lower continuum edge lies at the $X_{01}$ ground state energy
$-0.574E_0$. The density of $X_{01}$ states exhibits at this energy
an inverse square-root van Hove singularity  \cite{PRB2002}
typical for 2D excitations whose dispersion has an extremum
at a finite momentum.
The presence of the singularity may lead to peculiarities in
the $e^-$ scattering on the neutral exciton $X_{01}$ and one might
expect, in particular, formation of quasibound
three-particle states and optical resonances (see below).

Filled dots in Figs.~\ref{fig:t-levels},  \ref{fig:s-levels}
show positions of the excited bound three-particle states, denoted as
$(2e)$-$h$,
in which the hole is bound to the two-electron pair.
These states originate from the excited states of two electrons that
are always bound in 2D because of the confining effect of the magnetic
field. \cite{FQHE} These truly bound three-particle states form
discrete spectra and lie {\em above\/} free LL's.
They appear with relatively large positive values of $M_z$,
when the hole can be at a sufficiently large distance from the
two-electron pair. These states are not relevant for the considered
optical transitions and will not be discussed further.

There are also {\em many\/} bound states of charged triplet $X^-_{t01}$
and singlet $X^-_{s01}$ magnetoexcitons in the first hole LL lying below
the exciton continuum (open dots in
Figs.~\ref{fig:t-levels},  \ref{fig:s-levels}).
This is in contrast to the situation in the first electron and
zero hole LL's, ($N_eN_h$)=(10), where only one bound state,  the
strongly bound triplet $X_{t10}$, exists in the 2D high-field
limit. \cite{PRL2000}
This difference is explained by the fact that in ($N_eN_h$)=(10)
LL's a strongly bound neutral magnetoexciton  $X_{00}$ is
formed. As a result, the three-particle continuum of $X_{00}+ e^-_1$ states
is rather wide (width $E_0$),
and essentially one low-lying bound triplet state $X^-_{t10}$ can only
be supported outside the continuum.

\begin{figure}
\includegraphics[angle=-90,scale=0.33]{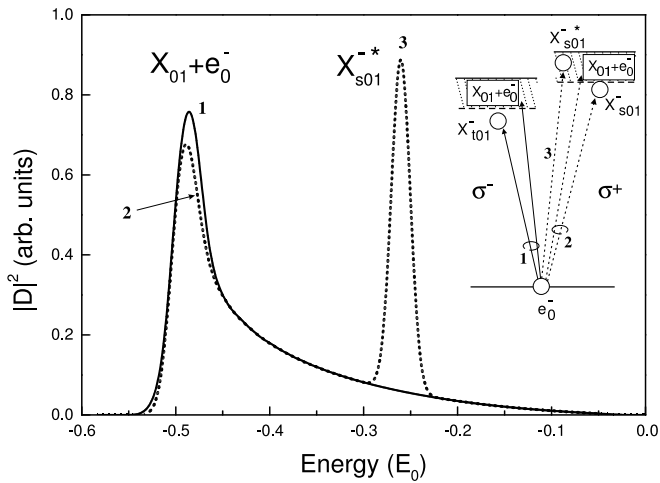}
\caption{       \label{fig:tran}
Dipole matrix elements of the
hole-ExCR transitions  in  $\sigma^{-}$ (solid line) and
$\sigma^+$ (dotted line)
polarizations at $\gamma = E_0/\hbar\omega_{\rm ch}= 1$.
Energy is counted from  $E_g(B) + \hbar\omega_{\rm ch}/2$.
Labeling of the peaks is explained in the inset,
see also  text and
Figs.~\protect\ref{fig:t-levels},\,\protect\ref{fig:s-levels}.}
\end{figure}

According to selection rule Eq.~(\ref{D}), the states that can optically be
excited from zero electron LL $N_e=0$ must have $M_z=0$. There is
one triplet and one singlet such states in the ($N_eN_h$)=(01)
LL's. These ExCR-active states are denoted by the filled triangles
in Figs.~3,\,4. The binding energies of the ExCR-active singlet
$X^-_{s01}$ and triplet $X^-_{t01}$ states are
$0.009E_0$ and $0.024E_0$, respectively.
(Note that these are not the ground states
and there are other low-lying triplet $X^-_{t01}$ and singlet $X^-_{s01}$
states that have larger binding energies.)
Therefore, the ExCR transitions in the 2DEG to the first hole LL
can terminate both in the {\em bound\/} states of charged magnetoexcitons,
$ e^-_0 + {\sl photon} \rightarrow X^-_{s(t)01}$, and, also, in the
continuum of scattering $X_{01}+ e^-_0$ states (see inset to
Fig.~\ref{fig:tran}).
This is in contrast to the situation with transitions to
the first electron LL, where transitions to the continuum are only
allowed. \cite{PRB2001}

\begin{figure}[b]
\includegraphics[angle=-90,scale=0.41]{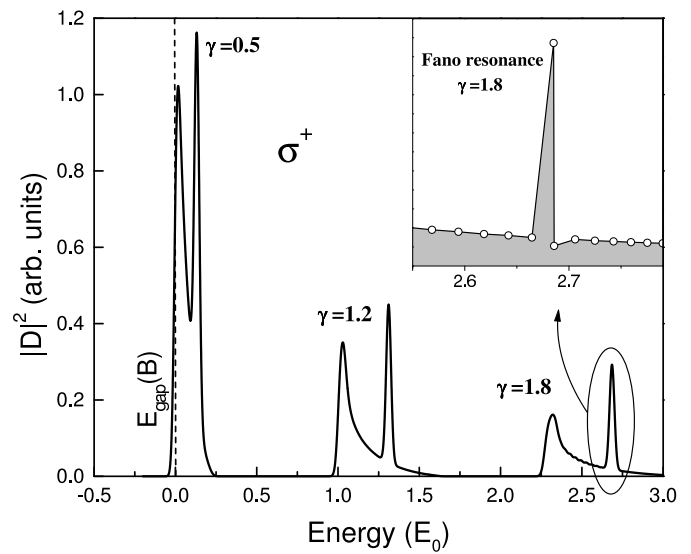}
\caption{       \label{fig:tran2}
The hole-ExCR spectra in  the $\sigma^{+}$  polarization at
three different magnetic fields
characterized by the dimensionless parameter
$\protect{\gamma = E_0/\hbar\omega_{\rm ch}}$.
Energy is counted from the fundamental band gap $E_g(B)$.
Inset shows the data with no Gaussian convolution for the resonance at
$\gamma = 1.8$.
}
\end{figure}

The calculated dipole matrix elements of the hole-ExCR transitions
in two circular polarizations $\sigma^{+}$ and $\sigma^{-}$ that
terminate, respectively, in the final singlet
$\Psi_{M_z=0 s \nu}^{(01)-}$ and triplet $\Psi_{M_z=0 t \nu}^{(01)-}$
states are shown in Fig.~\ref{fig:tran} for a representative case
$\gamma =1$. The dimensionless parameter
$\gamma = E_0/\hbar\omega_{\rm ch} \sim B^{1/2}$
characterizes the magnetic field strength.
The $\sigma^{+}$ spectra at three other magnetic fields are shown
in Fig.~\ref{fig:tran2}.
All energies are given in units of (the magnetic-field dependent value) $E_0$
taken at $\gamma=1$.  The spectra have been convoluted with a Gaussian of the
$0.015E_0$ width that mimics inhomogeneous broadening.
Note that both figures depict only the Coulomb interaction energies.
The shown ExCR transitions require the extra photon energy
$\hbar\omega_{\rm ch}$ (blue shifted) relative to the fundamental
band-gap absorption
$E_{\rm gap}(B)= E_g(0) + \hbar(\omega_{\rm ce} + \omega_{\rm ch})/2$
(with final states in the lowest LL's) and they are also
red-shifted $\simeq 0.5E_0 \sim B^{1/2}$ due to the Coulomb interactions.
The ExCR lines are also split in $\sigma^{+}$ and $\sigma^{-}$
polarizations by the Zeeman energies
$\mu_{B}(g_e S_{\rm ez} + g_h S_{\rm hz})B \sim B$.
The latter is schematically shown in the inset
to Fig.~\ref{fig:tran}.

It should be stressed that the ExCR transitions only arise because of LL
mixing. \cite{PRB2001}  The dependence of the
dipole transition matrix elements on magnetic field
is approximately $|D|^2 \sim B^{-1}$ (Fig.~\ref{fig:tran2}), as expected.
The {\em intensity} of transitions depends also on electron filling factor
$\nu_e$ (see above). Therefore, the hole-ExCR transitions are
suppressed in strong fields as
$\nu_e |D|^2 \sim n_e l_B^2 (E_0/\hbar\omega_{\rm ch})^2 \sim B^{-2}$.

The ExCR-active trions $X^-_{s01}$ and $X^-_{t01}$
have rather small binding energies.
Because of this, transitions to these bound states merge,
in the presence of moderate broadening, with the strong transitions
that terminate near the
lower continuum edge. These are the main peaks in
the hole-ExCR spectra (transitions~1 and 2 in Fig.~\ref{fig:tran}).
These peaks have
intrinsic finite linewidths, in high fields $\sim 0.1 E_0$, and
have asymmetric lineshapes with high-energy tails. This is because
the three-particle continuum is mainly responsible for these
transitions. Generally, the hole-ExCR absorption band becomes
wider and weaker with increasing magnetic field (Fig.~\ref{fig:tran2}).

The present theory predicts in high fields another
strong hole-ExCR feature, the prominent and rather {\em narrow\/}
higher-lying peak that can only be observed in the $\sigma^+$ polarization
(transition~3 in Fig.~\ref{fig:tran} and second peak in Fig.~\ref{fig:tran2}).
This resonance does not show any appreciable change in its width
with increasing $B$.
This peak might be associated with formation of a quasi-bound
three-particle resonance $X^{-*}_{s01}$, which lies
{\em within\/} the three-particle continuum
(see the inset to Fig.~\ref{fig:tran}).
The existence of the $2e$-$h$ resonances is physically plausible
in higher LL's because of the van Hove singularities in the density
of states of the neutral 2D mangnetoexcitons (see above).
A probability amplitude of finding all three particles in the same
region of real space is large for a well-defined resonance.
This may give large $e$-$h$ overlap \cite{PRB2001}
and lead to unexpectedly large oscillator strengths
of ExCR transitions to the continuum originating
from optically {\em inactive\/}  neutral
magnetoexcitons $X_{N_eN_h}$ ($N_e \neq N_h$).

The inset to Fig.~\ref{fig:tran2} presents the data with no Gaussian
convolution and shows some asymmetry in the shape of the optical resonance.
This is due to the interaction between the quasi-discrete level and the
continuum and is typical for a Fano-resonance. In the considered case
of $2e$-$h$ complexes in the first hole LL,
the interaction with the continuum appears to be rather weak and
only leads to a slight asymmetry in the resonance shape
(inset to Fig.~\ref{fig:tran2}).
This should be contrasted with the situation in the first electron LL,
where the interaction is much stronger, and the quasi-discrete state
may show up as an anti-resonance in the intraband absorption spectra
(see Fig.~2 in Ref.~\onlinecite{PRB2002}).
Details of formation of the three-particle resonances and the
evolution of such features in the spectra with decreasing magnetic
field are not currently understood.

The polarization dependence of the spectra may be explained by
different $e$-$e$ correlations in the final
spin-singlet ($\sigma^+$ polarization) and
spin-triplet ($\sigma^-$ polarization) three-particle $2e$-$h$
states;
the former are characterized by much stronger
$e$-$e$ repulsion. In the case considered this results in
formation of the higher lying ExCR-active resonance
$X^{-*}_{s01}$. Note also redistribution of the oscillator strength to
the higher-energy peak~3 at the expense of the main peak~2
that occurs in $\sigma^+$ polarization (see Fig.~\ref{fig:tran}).

Note that one might expect some qualitative or even
quantitative similarities between
the spectra of the hole-ExCR {\em interband\/} transitions in the
low-density 2DEG and the spectra of {\em intraband\/} internal
transitions of isolated negative trions $X^-$.
This is because these transitions may have similar or exactly the
same {\em final\/} three-particle correlated states in higher LL's.
Consider, for example, internal transitions
from the ground singlet  $X^-_s$ state that
has $M_z=0$. \cite{trions,PRL2002,Whit97}
When transitions to the first hole LL are induced by a
FIR-photon of $\sigma^-$ polarization, the final states
are $\Psi_{M_z s \nu}^{(01)-}$ with $M_z=-1$.
Note that in contrast to interband hole-ExCR, which is suppressed in
strong magnetic fields, such internal transitions are hole-CR active, strong
and gain strength \cite{JETPL99} with increasing $B$.
This may be useful for experimental studies of the 2DEG using
optically-detected cyclotron resonance techniques. \cite{PRL2002}

\section{   \label{sec:SU} Relation to shake-ups in 2DHG}

In this section, magneto-optics of an interacting
2DHG is discussed on a qualitative level in the approximation
neglecting the complex valence band structure.
Generally, a difference in electron and hole effective masses determines different
slopes of electron and hole LL's and, at finite fields, leads to different
interparticle correlations and different magneto-optical spectra
in the interacting 2DEG and 2DHG.
\begin{figure}[t]
\includegraphics[scale=0.5]{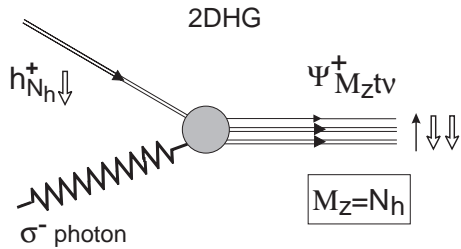}
\caption{   \label{fig:diag}
Interband absorption in the low-density spin-polarized 2DHG leading
to formation of charged ($2h$-$e$) three-particle states
$\Psi^+_{M_zt\nu}$ with $M_z = N_h$.
}
\end{figure}
In extremely strong magnetic fields
$\hbar\omega_{\rm ce}, \hbar\omega_{\rm ch} \gg  E_0$,
however, LL mixing is
suppressed and effective masses no longer determine interparticle
correlations. In this limit one can find a very simple relation between
the 2DEG and 2DHG magneto-optical spectra.

Consider, for example, a creation of an $e$-$h$ pair in the presence of
low-density spin-polarized 2DHG in zero LL $N_h=0$\,$\Downarrow$.
The exact optical selection rule involving the initial hole state in
$N_h$-th LL is  (Fig.~\ref{fig:diag})
 \begin{equation}
       \label{hExCR}
   D(\nu) = \langle \Psi^+_{M_z \, S_e S_h \nu}| \hat{\cal L}_{\rm PL}
              |h^{+}_{N_h} \rangle \sim \delta_{N_h,M_z} \, .
\end{equation}
Therefore, when the initial hole state belongs to $N_h=0$ LL,
the final positively charged $2h$-$e$ states
$\Psi^+_{M_z \, S_e S_h \nu}$
must have $M_z=0$. In $\sigma^-$ polarization these
are {\em hole\/} triplets ($\Downarrow$$\Downarrow$);
a short-hand notation $\Psi^+_{M_z=0 t \nu}$ will be used in what follows.
Notice now that the {\em negatively\/} charged electron triplet
state $\Psi^-_{M_z=0 t \nu}$
has---in the high field limit---exactly the same
Coulomb interaction energy (see Appendix~\ref{sec:App}).
Also, the state $\Psi^-_{M_z=0 t \nu}$ can be reached in transitions from
low-density spin-polarized 2DEG
in $N_e=0$\,$\uparrow$ LL  in same $\sigma^-$ polarization.
Therefore, the ExCR spectra
(and, more generally, absorption spectra)
of the spin-polarized ($\Downarrow$) 2DHG
can be obtained from the spin-polarized ($\uparrow$) 2DEG spectra
in the same polarization by simply
changing to the hole cyclotron energies
$\hbar\omega_{\rm ce} \rightarrow \hbar\omega_{\rm ch}$
and making the appropriate changes in the Zeeman energies.

At finite fields, LL mixing (especially for the hole)
breaks this electron-hole duality. In sufficiently strong
fields, however, such that $\hbar\omega_{\rm ch} \geq  E_0$,
one may expect the indicated quantitative similarities between the spectra
of the magnetically quantized 2DEG and 2DHG.

\section{\label{sec:X-} Shake-ups in magneto-PL of
                             isolated trions? Prohibited.}

As we have seen,  the ExCR (shake-up)
processes in magneto-photo{\em absorption\/}
are allowed in the low-density 2DEG and 2DHG.
Contrary to this,  the exact selection rules prohibit
shake-up processes in magneto-PL of {\em isolated\/} negative $X^-$
and positive $X^+$ charged trions.
Indeed, consider, for example, magneto-photoabsorption
in a low-density 2DEG with an isolated electron in
an arbitrary $N_e$-th LL in the initial state. According to the exact optical
selection rule
\begin{equation}
       \label{ExCrn}
   D(\nu) = \langle \Psi^-_{M_z \, S_e S_h \nu}| \hat{\cal L}_{\rm PL}
              |e^{-}_{N_e} \rangle  \sim \delta_{M_z,N_e}  \, ,
\end{equation}
the final negatively charged three-particle state $\Psi^-_{M_zS_e\nu}$
must have $M_z=N_e$ and may belong to {\em different\/} LL's;
this is generalization  \cite{PRL2000} of Eq.~(\ref{D}).

\begin{figure}[t]
\includegraphics[scale=0.5]{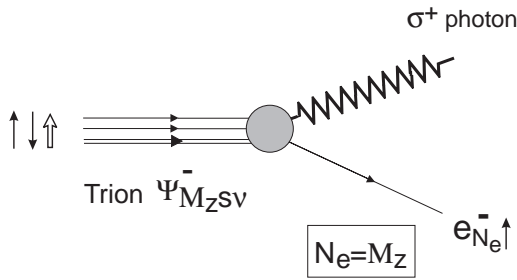}
\caption{\label{fig:diag2}
PL of an isolated negatively charged trion complex
$\Psi^-_{M_zs\nu}$.
Electron in the final state must reside
in the electron LL $N_e=M_z$.
}
\end{figure}

Consider now magneto-PL from an isolated charged trion state
$\Psi^-_{M_z \, S_e S_h \nu}$ that is not necessarily bound ($X^-$),
but may also belong to the continuum of scattering states ($X+e^-$).
The final state in magneto-PL is now a single electron (Fig.~\ref{fig:diag2}),
and the exact selection rule
\begin{equation}
       \label{X-SU}
   D(\nu)^* = \langle e^{-}_{N_e}  | \hat{\cal L}^{\dag}_{\rm PL}
   | \Psi^-_{M_z \, S_e S_h \nu} \rangle \sim \delta_{N_e,M_z} \,
\end{equation}
now states that the electron must reside in a {\em single  and given\/} LL
with number $N_e = M_z$.
Therefore,  {\em no various\/} LL's are achievable
in magneto-PL of a given trion state $\Psi^-_{M_zS_e\nu}$.
This result is easily extended to positively charged trions $X^+$
and holds at {\em arbitrary\/} fields in systems with
translational invariance and complex valence band structure
(in the axial approximation). \cite{PRL2000,PRB2002}

However,  the shake-up processes in magneto-PL attributed to
negatively \cite{Fink96}  $X^-$ and positively
\cite{GlasbergPRB2001} $X^+$ charged trions are commonly
observed in experiments in the {\em dilute\/} limit.
This may be interpreted as an indication toward the relevance of the various
scattering processes such as by disorder \cite{PRL2000,Ashkinadze}
and/or the remaining 2DEG for $X^-$ (or 2DHG for $X^+$).
It was suggested recently
\cite{Sanvitto} that the latter process, the scattering on
low-density 2DEG, may also give rise to appreciable oscillator strength of
the so-called ``dark'' triplet
$X^-$ state, \cite{Palacios96,PRL2000,Wojs2000} which
is also commonly observed in
experiments. \cite{trions,Yusa2001,Schueller2002,Broocks2002}

\section{\label{sec:summary} Summary}

This paper theoretically considered  hole-ExCR in a low-density
2DEG, an optical resonance in the 2DEG photoabsorption in which the hole
is excited to higher hole Landau levels.
This resonance may be observed in high magnetic fields as a fine structure
in the high-energy tail of the main magneto-photoabsorption peak of the 2DEG.
It has been shown, in particular,
that the high-field hole-ExCR has
different absorption patterns in two different circular
polarizations $\sigma^{\pm}$.
The duality between correlated states in the 2DEG and 2DHG
in the high magnetic field limit was indicated.
This may be useful for magneto-optical studies of
spin-dependent electron-hole correlations in
2D systems  in strong magnetic fields.


\begin{acknowledgments}
This work is supported in part by NSF grants DMR-0203560 and DMR-0224225
and by a research grant of CSUB.
\end{acknowledgments}


\appendix*
\section{ \label{sec:App} Duality between negatively and
             positively charged states}

In the high-field limit (when no LL mixing is allowed)
few-particle interacting states can be classified according to electron
and hole LL's ($N_eN_h$).
In this limit, the Coulomb {\em interaction\/}
energies of the positively charged ($2h$-$e$) hole triplet
states $\Psi^{(N_eN_h)+}_{M_z t \nu}$
and of the negatively charged ($2e$-$h$)
electron triplet states $\Psi^{(N_hN_e)-}_{-M_zt\nu}$
are the same; note the $M_z$ values and the order of LL labels.
This holds in the effective mass approximation neglecting the
complex valence band structure. In this approximation,
the orbital (envelope) wavefunctions of
the $2e$-$h$ and $2h$-$e$ states are related to each other by time reversal
(are complex conjugate to each other).  As a result, their Coulomb
energies coincide. \cite{JETPL97} The difference in their total energies
comes only from different cyclotron and spin Zeeman energies,
which can easily be taken into account.
The same, of course, holds for positively
$\Psi^{(N_eN_h)+}_{M_z s \nu}$
and negatively charged  $\Psi^{(N_hN_e)-}_{-M_z s \nu}$
singlet states whose interaction energies also coincide.

This duality between $2h$-$e$ and $2e$-$h$ states can be extended
in high fields to other (charged and neutral) complexes consisting of other
number of particles. \cite{JETPL97}
It is known, for example, that
dispersions of neutral magnetoexcitons $X_{N_eN_h}$ and $X_{N_hN_e}$
coincide in the high-field limit. \cite{L&L80}
Another closely related example is the duality between the 2D
correlated electron states that occur at
electron filling factors $\nu_e$ and $1-\nu_e$;
this result is known in the theory of the fractional
quantum Hall effect \cite{FQHE}
as the electron-hole symmetry in the lowest LL.


\end{document}